\documentclass[pra,showpacs,graphicx,multicols,amsmath,amssymb]{revtex4}

\usepackage{amssymb}
\usepackage[dvips]{color}
\usepackage{amssymb}
\usepackage{amsfonts}
\usepackage{amsmath}
\usepackage{graphicx}

\newcommand{\bfk}{{\bf k}}
\newcommand{\ek}{\epsilon_\bfk}
\newcommand{\delk}{\Delta_\bfk}
\newcommand{\psd}{\eta}
\newcommand{\cks}{c_{\bfk,\sigma}}
\newcommand{\aks}{a_{\sigma,\bfk}}
\newcommand{\cku}{c_{\bfk,\uparrow}}

\newcommand{\cmkd}{c_{-\bfk,\downarrow}}
\newcommand{\aku}{a_{1,\bfk}}

\newcommand{\amkd}{a_{2,-\bfk}}

\newcommand{\bz}{b}
\newcommand{\bx}{B_x}
\newcommand{\by}{B_y}

\newcommand{\jmk}{S_-^\bfk}
\newcommand{\jmkp}{S_-^{\bfk'}}
\newcommand{\jpkp}{S_+^{\bfk'}}
\newcommand{\jpk}{S_+^\bfk}
\newcommand{\jpmk}{S_\pm^\bfk}
\newcommand{\jxk}{S_x^\bfk}
\newcommand{\jyk}{S_y^\bfk}
\newcommand{\jzk}{S_z^\bfk}
\newcommand{\jik}{S_i^\bfk}
\newcommand{\jm}{J_-}
\newcommand{\jp}{J_+}
\newcommand{\jx}{J_x}
\newcommand{\jy}{J_y}
\newcommand{\jz}{J_z}
\newcommand{\ji}{J_i}
\newcommand{\jj}{J_j}
\newcommand{\jk}{J_k}
\newcommand{\cjm}{{\cal J}_-}
\newcommand{\cjp}{{\cal J}_+}
\newcommand{\cjx}{{\cal J}_x}
\newcommand{\cjy}{{\cal J}_y}
\newcommand{\cjz}{{\cal J}_z}
\newcommand{\cji}{{\cal J}_i}
\newcommand{\cjmk}{{\cal S}_-^\bfk}
\newcommand{\cjpk}{{\cal S}_+^\bfk}

\newcommand{\cjik}{{\cal S}_i^\bfk}
\newcommand{\kmk}{K_-^\bfk}

\newcommand{\kpk}{K_+^\bfk}
\newcommand{\kpmk}{K_\pm^\bfk}

\newcommand{\kzk}{K_z^\bfk}
\newcommand{\km}{K_-}
\newcommand{\kp}{K_+}
\newcommand{\kpm}{K_\pm}
\newcommand{\kx}{K_x}
\newcommand{\ky}{K_y}
\newcommand{\kz}{K_z}
\newcommand{\ki}{K_i}
\newcommand{\ckm}{{\cal K}_-}
\newcommand{\ckp}{{\cal K}_+}
\newcommand{\ckx}{{\cal K}_x}
\newcommand{\cky}{{\cal K}_y}
\newcommand{\ckz}{{\cal K}_z}
\newcommand{\cki}{{\cal K}_i}

\newcommand{\clz}{{\cal L}_z}

\begin{document}

\title{Matter-wave squeezing and the generation of $SU(1,1)$ and
$SU(2)$ coherent-states via Feshbach resonances}

\author{I. Tikhonenkov,  E. Pazy, Y. B. Band, and A. Vardi}
\affiliation{Department of Chemistry, Ben Gurion University of Negev, P.O.B. 653,
Beer Sheva 84105, Israel}

\begin{abstract}
Pair operators for boson and fermion atoms generate $SU(1,1)$ and
$SU(2)$ Lie algebras, respectively.  Consequently, the pairing of
boson and fermion atoms into diatomic molecules via Feshbach
resonances, produces $SU(1,1)$ and $SU(2)$ coherent states, making
bosonic pairing the matter-wave equivalent of parametric coupling and
fermion pairing equivalent to the Dicke model of quantum optics.  We
discuss the properties of atomic states generated in the dissociation
of molecular Bose-Einstein condensates into boson or fermion
constituent atoms.  The $SU(2)$ coherent states produced in
dissociation into fermions give Poissonian atom-number distributions,
whereas the $SU(1,1)$ states generated in dissociation into bosons
result in super-poissonian distributions, in analogy to two-photon
squeezed states.  In contrast, starting from an atomic gas produces
coherent number distributions for bosons and super-poissonian
distributions for fermions.
\end{abstract}


\maketitle


\section{Introduction}

The behavior of a gas of non-interacting particles close to the
absolute zero of temperature depends solely on their quantum
statistics.  Whereas fermions obey Pauli exclusion, manifested in the
equal-time anticommutation relations of their field operators, bosons
are subject to Bose enhancement, implicit in their field operator
commutators.  For interacting particles, the interaction affects the
pair-statistics of fermions and bosons.  Pairing models have attracted
renewed interest since Feshbach resonances \cite{Feshbach,Inouye98}
have been employed to realize molecular Bose-Einstein condensates with
fermionic \cite{Regal03, Greiner03, Jochim03, Zwierlein03} and bosonic
\cite{Claussen02, Herbig03, Durr04} constituent atoms, and the ensuing
research of the BEC-BCS crossover \cite{Regal04, Zwierlein04,
Bartenstein04, Bourdel04,Kinast04, Chin04, Greiner05, Zwierlein05}.
Recently, some attention was given to the relation between the quantum
statistics of the atomic gas in which Feshbach or optical association
is performed, and the resulting number-statistics of the atomic and
molecular fields generated in the process \cite{Miyakawa05,
MeiserPRA05, MeiserPRL05,Uys05}.  It was shown that, whereas the
molecular field produced in boson association will initially be in a
Glauber coherent state, $|\alpha \rangle$, defined such that $a
|\alpha \rangle = \alpha |\alpha \rangle$ where $a$ is a destruction
operator, having constant particle-number fluctuations [i.e., $(\Delta
n)^2/\langle n\rangle=1$, where $n=a^\dag a$ and $(\Delta n)^2=\langle
n^2\rangle-\langle n\rangle^2$], the corresponding number
distributions for fermions will be super-Poissonian \cite{MeiserPRL05}
with $(\Delta n)^2$ exceeding $\langle n\rangle$.  The coherence of
the boson-association field was attributed to collective association,
whereas the chaotic number distributions for fermion-association were
related to the individual association of fermionic atom pairs.

Here we explain and quantify these differences in the molecular
number-statistics in terms of the commutation relations of fermion and
boson {\it pair} operators.  It is well known that pair operators for
fermions and bosons generate $SU(2)$ and $SU(1,1)$ algebras,
respectively \cite{Wodkiewicz85,Andreev04,Barankov04}.  Consequently,
the atomic field produced in the dissociation of a molecular BEC into
fermion atoms will be in an $SU(2)$ coherent state with Poissonian
number distribution, whereas boson atoms thus generated, will be in an
$SU(1,1)$ coherent state, corresponding to a squeezed state of the
Wigner-Weyl algebra, with a super-Poissonian distribution.  Using the
simple mapping between $SU(2)$ and $SU(1,1)$ it is shown that for
association, $SU(1,1)$ coherent states will initially dominate fermion
pairing whereas $SU(2)$ states will be generated for bosons.  Boson
association (unlike boson dissociation) is not a collective effect
since the molecular field can be replaced by a macroscopic c-number,
rendering the initial molecule production process perfectly linear.
The super-Poissonian statistics of fermion association on the other
hand, actually result in from collective behavior of the fermionic
association.  In particular, it will also show up in the degenerate
fermionic case, where all atom pairs `emit' molecules in-phase.

In section \ref{sec_Dynam_Eqs} we discuss the dynamical equations for
associative pairing of fermionic and bosonic atoms, Sec.~\ref{sect_cs}
describes the bosonic and fermionic coherent states of the $SU(1,1)$
and $SU(2)$ algebras generated by the angular momentum like operators
respectively, Sec.~\ref{sect_short_time} discusses the short time
dynamics of the dissociation of a molecular BEC into bosonic and
fermionic atoms, and Sec.~\ref{sect_summary} concludes the paper.

\section{Dynamical Equations}  \label{sec_Dynam_Eqs}

We begin by considering the dynamical equations for associative
pairing of fermionic and bosonic atoms, highlighting similarities and
differences resulting in from the underlying pair statistics of
$SU(2)$ for fermions and of $SU(1,1)$ for bosons.  As shown below,
when the atomic motion is slow with respect to other timescales, the
atom-molecule pairing Hamiltonians map onto two quantum-optical
paradigmatic systems; The pairing of fermion atoms is a matter-wave
equivalent of Dicke superradiance \cite{Dicke54}, whereas the
dissociation into bosonic atoms is analogous to parametric
downconversion \cite{Walls_Milburn}.
 
\subsection{Fermion atoms - $SU(2)$ algebra}

Consider the single molecular mode association/dissociation
Hamiltonian
\begin{equation}
H=\sum_{\bfk,\sigma} \ek\cks^\dag\cks+{\cal E}\bz^\dag\bz +g \left( \!
\bz^\dag\sum_{\bf k} \cku\cmkd + h.c. \!  \right)~,
\label{ham}
\end{equation}
where $\ek=\hbar^2k^2/2m$ is the kinetic energy of an atom with mass
$m$, $\cal E$ is the molecular energy, containing kinetic and binding
contributions, and $g$ is the atom-molecule coupling strength.  The
annihilation operators for the atoms, $\cks$, obey fermionic
anticommutation relations, whereas the molecular annihilation operator
$\bz$ obeys a bosonic commutation relation.  The single mode
approximation is justified when the molecular dispersion due to the
presence of a molecular momentum spread, is slow with respect to any
other timescale in the problem.  It becomes exact for a molecular BEC,
when molecular translation is completely frozen.  For simplicity, we
have also omitted background non-reactive atom-atom scattering.  As
will be evident from the discussion below, these interactions can be
easily incorporated, as long as they are dominated by $({\bf
k}\uparrow,-{\bf k}\downarrow)$ pairing.  While this assumption is
well-justified for fermions in the BCS state, it is a gross
oversimplification for bosons.  We thus expect that our results will
be restricted to the case where background open-channel interactions
are small with respect to closed-channel atom-molecule coupling, i.e.,
to narrow Feshbach resonances.

For fermionic atomic field operators, the model Hamiltonian
(\ref{ham}) can be written using only the atomic $SU(2)$ generators
\cite{Andreev04,Barankov04}
\begin{equation}
\jmk=\cku\cmkd~,~~
\jpk=\cmkd^\dag\cku^\dag~,~~
\jzk={1\over 2}\left(-1+\cku^\dag\cku+\cmkd^\dag\cmkd\right),
\label{jkdef}
\end{equation}
obeying the canonical angular-momentum commutation relations
\begin{equation}
[\jpk,\jmk]=2\jzk~,~~[\jzk,\jpmk]=\pm\jpmk~.
\label{sutcom}
\end{equation}
Using Eqs. (\ref{jkdef}), Hamiltonian (\ref{ham}) may be rewritten
as
\begin{equation}
H=\sum_{\bfk} \ek\left(2\jzk+1\right)+{\cal
E}\bz^\dag\bz+g\sum_{\bfk}\left(\bz^\dag\jmk+\jpk\bz\right),
\label{hamsut}
\end{equation}
resulting in the Heisenberg equations of motion,
\begin{eqnarray}
i\dot\jpk&=&[\jpk,H]=-2\ek\jpk+2g\bz^\dag\jzk,\nonumber\\
i\dot\jmk&=&[\jmk,H]=2\ek\jmk-2g\jzk\bz,\nonumber\\
i\dot\jzk&=&[\jzk,H]=g\left(\jpk\bz-\bz^\dag\jmk\right),\nonumber\\
\label{ferone}
i\dot\bz&=&[\bz,H]={\cal E}\bz+g\sum_\bfk\jmk~.
\end{eqnarray}
Defining the Hermitian operators
\begin{equation}
\jxk=\frac{\jpk+\jmk}{2}~,\jyk=\frac{\jpk-\jmk}{2i}~,
\bx=\frac{\bz+\bz^\dag}{2}~,\by=\frac{\bz-\bz^\dag}{2i}~,
\end{equation}
equations (\ref{ferone}) transform into:
\begin{eqnarray}
\dot\jxk=-2\ek\jyk-2g\by\jzk,&~~\dot\jyk=2\ek\jxk-2g\bx\jzk,
&~~\dot\jzk=2g\left(\bx\jyk+\by\jxk\right),\nonumber\\
\label{fertwo}
\dot\bx={\cal E}\by-g\sum_\bfk\jyk,&~~\dot\by=-{\cal
E}\bx-g\sum_\bfk\jxk&~.
\end{eqnarray}
System (\ref{fertwo}) satisfies the conservation of the individual
spin angular momenta, with the $SU(2)$ Casimir operators
\begin{equation}
{{\bf S}^{\bfk}}^2=\jzk(\jzk-1)+\jpk\jmk=s(s+1)~,
\end{equation}
with $s=1/2$, as well as total number conservation  
\begin{equation}
\left(\bx^2+\by^2-1/2\right)+\sum_{\bfk}(\jzk+1/2)=N/2~,
\end{equation}
where $N=2\bz^\dag\bz+\sum_{\bfk,\sigma}\cks^\dag\cks$. Defining,
\begin{equation}
\cjpk=\jpk\bz~,~\cjmk=\bz^\dag\jmk~,
\end{equation} 
the dynamical equations (\ref{ferone}) take the form
\begin{eqnarray}
i\dot\cjpk&=&\delk\cjpk+2g\bz^\dag\bz\jzk+g\sum_{\bfk'}\jpk\jmkp~,\nonumber\\
i\dot\cjmk&=&-\delk\cjmk-2g\bz^\dag\bz\jzk-g\sum_{\bfk'}\jpkp\jmk~,\nonumber\\
\label{ferthree}
i\dot\jzk&=&g\left(\cjpk-\cjmk\right)~,
\end{eqnarray}
with $\delk={\cal E}-2\ek$.

For the degenerate case, $\ek \rightarrow \epsilon$, which yields
\begin{equation}
\delk \rightarrow \Delta={\cal E}-2\epsilon ~.
\end{equation} 
Hamiltonian (\ref{hamsut}) is, up to an insignificant $c$-number
shift, just the Dicke Hamiltonian \cite{Dicke54} and the dynamical
equations become
\begin{eqnarray}
i\dot\cjp&=&\Delta\cjp+2g\bz^\dag\bz\jz+g\jp\jm~,\nonumber\\
i\dot\cjm&=&-\Delta\cjm-2g\bz^\dag\bz\jz-g\jp\jm~,\nonumber\\
\label{ferfour}
i\dot\jz&=&g\left(\cjp-\cjm\right)~,
\end{eqnarray}  
where $J_i=\sum_\bfk \jik~,{\cal J}_i=\sum_\bfk\cjik$. In order to
get a closed set of equations for $\cjp,\cjm$ and $\jz$, we use the
$SU(2)$ Casimir operator,
\begin{equation}
\jz(\jz-1)+\jp\jm=j(j+1) ~,
\label{cassut}
\end{equation}
and number conservation
\begin{equation}
\bz^\dag\bz=\frac{N}{2}-\left(j+\jz\right) ~,
\label{nconserv}
\end{equation}
where $N=2\bz^\dag\bz+\sum_{\bfk,\sigma}\cks^\dag\cks$ is the total
number of atoms and $j$ is the number of available energy levels
(containing at most $4j$ particles, because each level can accommodate
$\bfk\uparrow,\bfk\downarrow,-\bfk\uparrow,-\bfk\downarrow$ atoms).
Substituting Eq.~(\ref{cassut}) and Eq.~(\ref{nconserv}) into
Eqs.~(\ref{ferfour}), we obtain
\begin{eqnarray}
i\dot\cjp&=&\Delta\cjp-g\left[3\jz^2-(N-2j)\jz-j^2-j-\jz\right]~,\nonumber\\
i\dot\cjm&=&-\Delta\cjm+g\left[3\jz^2-(N-2j)\jz-j^2-j-\jz\right]~,\nonumber\\
\label{ferfive}
i\dot\jz&=&g\left(\cjp-\cjm\right)~.
\end{eqnarray}  
Defining the normalized operators
\begin{equation}
\cjx=\frac{{\cjp+\cjm}}{2(N/4)^{3/2}}~,~\cjy=\frac{{\cjp-\cjm}}{{2i(N/4)^{3/2}}}~,
~\cjz={\jz\over{N/4}}~,
\end{equation}
we finally obtain the dynamical equations
\begin{eqnarray}
\label{cjxdot}
\dot\cjx&=&\Delta\cjy~,\\
\label{cjydot}
\dot\cjy&=&-\Delta\cjx+\frac{g\sqrt{N}}{2}\left[3\cjz^2-\left(4-2\psd\right
)\cjz-\psd^2\right]-{{2g}\over \sqrt{N}}(\psd+\cjz)~,\\
\label{fersix}
\dot\cjz&=&g\sqrt{N}\cjy~,
\end{eqnarray}   
where $\psd=4j/N$ denotes the number of quantum states per particle or
the inverse phase space density. For a thermal gas $\psd\gg 1$,
whereas for a Fermi degenerate gas $\psd=1$. For a filled Fermi sea, 
$\psd$ attains its minimal value of unity, and Eq.~(\ref{cjydot}) 
can be replaced by
\begin{equation}
\dot\cjy=-\Delta\cjx-\frac{g\sqrt{N}}{2}\left[(1-\cjz)(1+3\cjz)\right]
-{2g\over \sqrt{N}}(1+\cjz)~.\\
\end{equation}

\subsection{Boson Atoms - $SU(1,1)$ algebra}

We next consider the coupling of a molecular BEC into {\it bosonic}
atom pairs. The single molecular mode Hamiltonian reads,
\begin{equation}
H=\sum_{\bfk,\sigma} \ek\aks^\dag\aks+{\cal E}\bz^\dag\bz + g \left(\!
\bz^\dag\sum_{\bf k} \aku\amkd + h.c. \!  \right)~,
\label{hambos}
\end{equation}
where $\ek$, $\cal E$, $g$, and $\bz$ have the same meaning as in
Eq.~(\ref{ham}) and the atomic annihilation operators $\aks$, denoting
two atom species, now obey bosonic commutation relations. 

The pertinent algebra for bosonic atom operators is
$SU(1,1)$, because the commutator of  $\kmk=\aku\amkd$ and 
$\kpk=\amkd^\dag\aku^\dag$ is
\begin{equation}
[\kpk,\kmk]=\left(-1-\aku^\dag\aku-\amkd^\dag\amkd\right)\equiv -2\kzk
~,
\end{equation}
so that the three generators $\kpk,\kmk,\kzk$ obey $SU(1,1)$
commutation relations
\begin{equation}
[\kpk,\kmk]=-2\kzk~,~[\kzk,\kpmk]=\pm\kpmk~,
\label{suoocom}
\end{equation}
differing only in the sign of $[\kpk,\kmk]$ from the commutation
relation between the $SU(2)$ generators, stipulated in
Eq.~(\ref{sutcom}).  Hamiltonian (\ref{hamsut}) is thus replaced by
the $SU(1,1)$ Hamiltonian,
\begin{equation}
H=\sum_{\bfk}\ek\left(2\kzk-1\right)+{\cal
E}\bz^\dag\bz+g\sum_{\bfk}\left(\bz^\dag\kmk+\kpk\bz\right),
\label{hamsuoo}
\end{equation} 
leading to the Heisenberg equations of motion,
\begin{eqnarray}
i\dot\kpk&=&[\kpk,H]=-2\ek\kpk-2g\bz^\dag\kzk ~, \nonumber \\
i\dot\kmk&=&[\kmk,H]=2\ek\kmk+2g\kzk\bz ~, \nonumber \\
i\dot\kzk&=&[\kzk,H]=g\left(\kpk\bz-\bz^\dag\kmk\right)~ , \nonumber \\
\label{bosone}
i\dot\bz&=&[\bz,H]={\cal E}\bz+g\sum_\bfk\kmk ~.
\end{eqnarray}

For degenerate atomic energy levels, the boson Hamiltonian
(\ref{hamsuoo}) is identical to the model Hamiltonian of parametric
downconversion \cite{Walls_Milburn}.  Following the same procedure as
in the previous section, we obtain for boson degenerate modes,
\begin{eqnarray}
i\dot\ckp&=&\Delta\ckp-2g\bz^\dag\bz\kz+g\kp\km~,\nonumber\\
i\dot\ckm&=&-\Delta\ckm+2g\bz^\dag\bz\kz-g\kp\km~,\nonumber\\
\label{bosthree}
i\dot\kz&=&g\left(\ckp-\ckm\right)~,
\end{eqnarray} 
where $\kpm=\sum_\bfk \kpmk$, $\kz=\sum_\bfk\kzk$, $\ckp=\kp\bz$, and
$\ckm=\bz^\dag\km$. In contrast to the unitary $SU(2)$ case where we 
had 
$$-j\le\langle \jz \rangle\le -j+\min\{N/2,2j\},$$ 
reducing to $-j\le\langle \jz\rangle\le j$ for $N=4j$, we now have 
$$k\le\langle\kz\rangle\le k+N/2,$$ 
where $4k$ denotes the number of boson atomic modes. However, we can
still 
eliminate $\kpm$ and $\bz$ by using 
number conservation and the $SU(1,1)$ Casimir operator,
\begin{equation}
\bz^\dag\bz={N\over 2}-(\kz-k)~,
\end{equation}
\begin{equation}
\kz(\kz-1)-\kp\km=k(k-1)~,
\label{suoocas}
\end{equation}
resulting in the dynamical equations
\begin{eqnarray}
i\dot\ckp&=&\Delta\ckp+g\left[3\kz^2-(N+2k)\kz-k^2+k-\kz\right]~,\nonumber\\
i\dot\ckm&=&-\Delta\ckm-g\left[3\kz^2-(N+2k)\kz-k^2+k-\kz\right]~,\nonumber\\
\label{ferfour''}
i\dot\kz&=&g\left(\ckp-\ckm\right)~.
\end{eqnarray}
We define, as we did for fermion atoms,
\begin{equation}
\ckx={{\ckp+\ckm}\over{2(N/4)^{3/2}}}~,~\cky={{\ckp-\ckm}\over{2i(N/4)^{3/2}}}~,
~\ckz={\kz\over
(N/4)}~,
\end{equation}
and using these definitions, the dynamical equations are transformed
to the final form
\begin{eqnarray}
\dot\ckx&=&\Delta\cky~,\nonumber\\
\label{ckydot}
\dot\cky&=&-\Delta\ckx-\frac{g\sqrt{N}}{2}\left[3\ckz^2-\left(4+2\psd
\right)\ckz-\psd^2\right]+{2g\over \sqrt{N}}(\ckz-\psd)~,\nonumber\\
\label{bosfive}
\dot\ckz&=&{g\sqrt{N}}\cky~.
\end{eqnarray}

In order to gain better insight on the relation between the fermion
equation (\ref{fersix}) and the boson equation (\ref{bosfive}) we
define the number difference operator $\clz=(2n_b-n_a)/N=1+\psd-\ckz$,
whose expectation value, like the expectation value of $\cjz$,
corresponds to the atom-molecule population imbalance.  With this
definition, we have
\begin{equation}
3\ckz^2-\left(4+2\psd \right)\ckz-\psd^2=3\clz^2-\left(2+4\psd
\right)\clz-1~,
\end{equation}
\begin{equation}
\ckz-\psd=1-\clz~,
\end{equation}
and the dynamical equations (\ref{bosfive}) assume the form
\begin{eqnarray}
\dot\ckx&=&\Delta\cky~,\nonumber\\
\label{ckydot'}
\dot\cky&=&-\Delta\ckx-\frac{g\sqrt{N}}{2}\left(3\clz+1\right)\left(\clz-1\right)+{2g\over
\sqrt{N}}\left[(4k-1)\clz+1\right]~,\nonumber\\
\label{bossix}
\dot\clz&=&-g\sqrt{N}\cky~.
\end{eqnarray}

The two atomic-modes case with Hamiltonian
\begin{equation}
H=\epsilon \left(a_1^\dag a_1 + a_2^\dag a_2\right)+
{\cal E}\bz^\dag\bz+g\left(\bz^\dag a_1a_2+a_2^\dag a_1^\dag\bz\right)
\end{equation}
is obtained from Eqs.~(\ref{bossix}) by substituting $k=1/2$ (because
the minimum value of $\kz$, obtained where no atoms are present, is
$1/2$).  It is easily verified that the resulting equations of motion
for $\ckx,\cky,\clz$ are identical up to the sign of $g$, with the
fermion equations for $\cjx,\cjy,\cjz$ when $\psd=1$.  Noting that
$\cjz$ for $\eta=1$ and $\clz$ have inverse interpretation (i.e. the
former equals $(n_a-2n_b)/N$ and the latter is $(2n_b-n_a)/N$) we see
that the dynamics of degenerate fermion association maps into two-mode
boson dissociation and vice versa.

\section{Coherent States}  \label{sect_cs}

Having developed the time-dependent many-body formalism and
established the connection with the quantum-optical paradigms, we turn
to the investigation of the dissociation of a molecular BEC consisting
either of fermionic or bosonic constituent atoms.  For sufficiently
short times, we neglect molecular fluctuations and treat the molecular
field $\bz$ as an undepleted pump, replacing it by the $c$-number
$\sqrt{N/2}$.  The resulting Hamiltonian for fermion (boson) atoms
under this approximation, thus consists of linear sums of operators
generating the $SU(2)$ ($SU(1,1)$) algebra.  Consequently, generalized
coherent matter states of the pertinent Lie algebras \cite{Wodkiewicz85}, 
can be dynamically generated in the dissociation of molecular BECs. In this
section we briefly discuss the properties of $SU(2)$ ($SU(1,1)$)
coherent states generated in the dissociation of a molecular BEC into
fermion (boson) atoms.

\subsection{$SU(2)$ Coherent states}

The generalized coherent states associated with the unitary
representations of the $SU(2)$ Lie algebra, are parametrized by the
two polar (Euler) angles $\theta$ and $\phi$ corresponding to
rotations of the fully stretched atomic vacuum state $|j,-j\rangle$
(where $|j,m\rangle$ denote the usual mutual eigenstates of the
Casimir operator ${\bf J}^2$ and of the number difference operator
$\jz$, i.e., ${\bf J}^2|j,m\rangle=j(j+1)|j,m\rangle$,
$\jz|j,m\rangle=m|j,m\rangle$ with $m=-j,\dots,j$) about the $\jx$
and $\jz$ axes, respectively:
\begin{equation}  \label{coherento}
|\theta,\phi\rangle \equiv
\exp\left(-i\phi\jz\right)\exp\left(-i\theta\jx\right) |j,-j\rangle =
\exp\left(\alpha \jp - \alpha^* \jm\right) |j,-j\rangle~,
\end{equation}  
with $\alpha=(\theta/2)\exp(-i\phi)$ \cite{SU2_coherent}.  Definition
(\ref{coherento}) results in the familiar expansion of $SU(2)$
coherent states in terms of number (Fock) states
\begin{eqnarray}
|\theta,\phi\rangle&=&\left[1+\tan^2\left({\theta\over
2}\right)\right]^{-j}\sum_{m=-j}^j\left[\tan\left({\theta\over
2}\right)\exp(-i\phi)\right]^{j+m}\left(
\begin{array}{c} 
2j\\
j+m
\end{array}\right)^{1/2}|j,m\rangle\ ~.
\label{coherentt}
\end{eqnarray}

\begin{figure}
\centering
\includegraphics[scale=0.7]{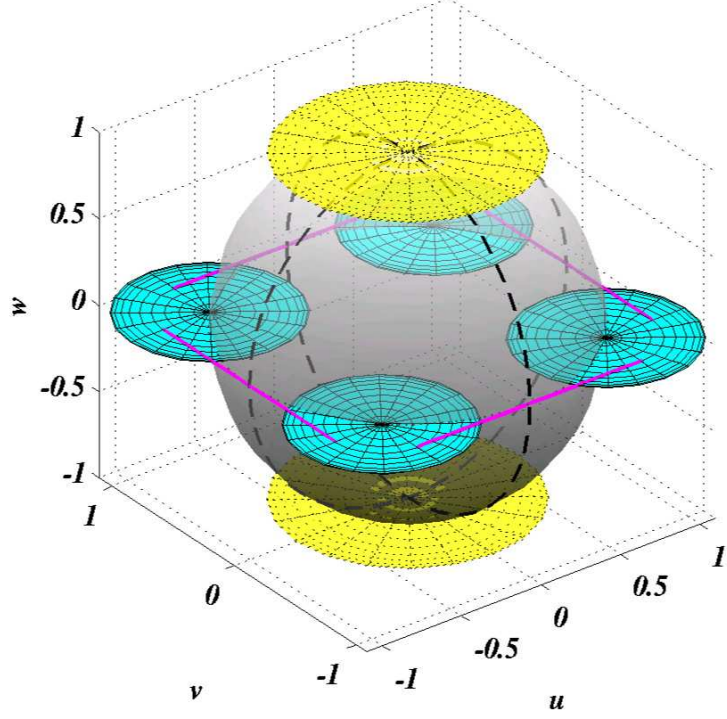}
\caption{Bloch sphere (shaded shell) and coherent states of $SU(2)$.
Dashed black curves mark intelligent coherent states.  Ellipses denote
$\Delta\jx$ and $\Delta\jy$ variance for ten coherent states: the
atomic and molecular vacuum states (yellow), four non-intelligent
states (cyan) and the four squeezed, minimum uncertainty states
(magenta).}
\label{fig1}
\end{figure}

Using either Eq.~(\ref{coherento}) or Eq.~(\ref{coherentt}) it is
easily verified that 
\begin{eqnarray}
\label{jxexp}
\langle\jx\rangle&=&j\sin\theta\cos\phi,\\
\label{jyexp}
\langle\jy\rangle&=&j\sin\theta\sin\phi,\\ 
\label{jzexp}
\langle\jz\rangle&=&j\cos\theta, 
\end{eqnarray}
so that the expectation values of $\bf J$ are restricted
to the Bloch sphere of radius $j$, as depicted in Fig.~\ref{fig1}. 
The coherent state variance of these operators is
\begin{eqnarray}
\label{djx}
\Delta\jx^2&=&{j\over 2}\left(1-\sin^2\theta\cos^2\phi\right) ~, \\
\label{djy}
\Delta\jy^2&=&{j\over 2}\left(1-\sin^2\theta\sin^2\phi\right) ~, \\
\label{djz}
\Delta\jz^2&=&{j\over 2}\sin^2\theta ~.
\end{eqnarray}
The total variance of coherent states is thus also bound because
$|\Delta {\bf J}|^2=\langle{\bf J}^2\rangle-\langle{\bf
J}\rangle^2=j(j+1)-j^2=j$. The 
commutation relations (\ref{sutcom}), lead to the uncertainty
relations 
\begin{equation}
\Delta\ji\Delta\jj\ge {1\over 2}|c_{ij}^k\langle \jk\rangle|,
\label{sutuncert}
\end{equation}
where $c_{ij}^k=\epsilon_{ij}^k$ are the $SU(2)$ structure constants.
In particular, for $\jx$ and $\jy$ we have 
\begin{equation}
\Delta\jx\Delta\jy\ge {1\over 2}|\langle \jz\rangle|,
\label{jxjyuncert}
\end{equation}
In Fig.~\ref{fig1} we plot the expectation values of ${\bf
J}/j=(u,v,w)$ for $SU(2)$ coherent states, as well as the $\Delta \jx$
and $\Delta \jy$ variance of ten such states.  Coherent states for
which inequality (\ref{jxjyuncert}) is an equality are referred to as
'intelligent states' or 'ideal coherent states'.  
From Eqs.~(\ref{djx}), (\ref{djy}), and
(\ref{jzexp}) we obtain that $SU(2)$ intelligent states are found for
$\phi=0,\pi/2,\pi,3\pi/2$ and arbitrary $\theta$, as depicted by
dashed curves in Fig.~\ref{fig1}.  A subset of the intelligent states
are the minimum uncertainty states with $\phi=0,\pi/2,\pi,3\pi/2$ and
$\theta=\pi/2$ (denoted by magenta ellipsoids in Fig.~\ref{fig1}), for
which the r.h.s. of Eq.~(\ref{jxjyuncert}) is minimized, with
$\Delta\jx\Delta\jy=0$.  While the states with $\theta=0$ and $\phi$
arbitrary (yellow disks) are also intelligent, their value of
$\Delta\jx\Delta\jy=j/2$ is in fact maximal and larger than
$\Delta\jx\Delta\jy=j/4$ obtained for the non-intelligent states
denoted by cyan disks.

\begin{figure}
\centering
\includegraphics[scale=0.7]{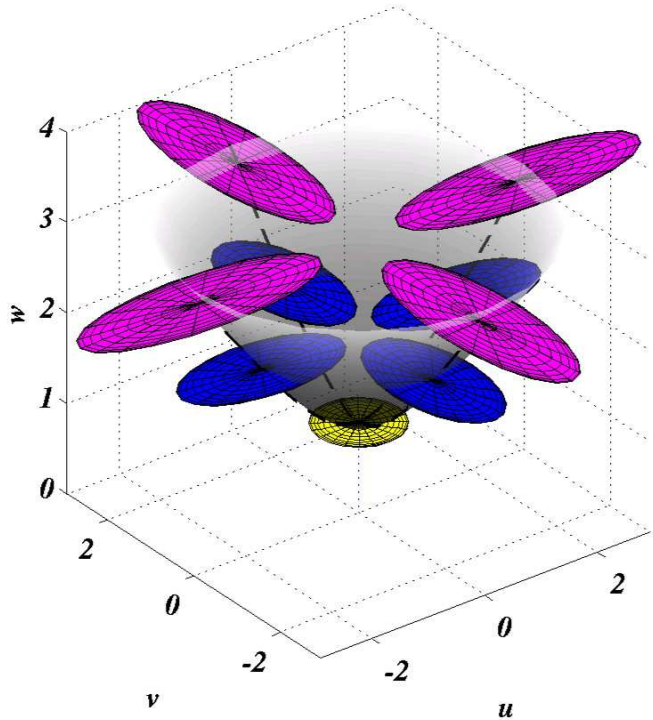}
\caption{Surface of motion of $\langle{\bf K}\rangle=(u,v,w)$ (shaded
paraboloid) and coherent states of $SU(1,1)$.  Dashed black curves
mark intelligent coherent states.  Ellipses denote the $\Delta\kx$ and
$\Delta\ky$ variances for nine intelligent coherent states.  Whereas
the atomic vacuum (yellow ellipse) is a minimum uncertainty state with
equal variance in the $\ckx$ and $\cky$ directions, other intelligent
states (magenta,blue) are squeezed.}
\label{fig2}
\end{figure}

\subsection{SU(1,1) Coherent states}

The mutual eigenstates of the $SU(1,1)$ Casimir operator
(\ref{suoocas}) and of $\kz$ form the basis set:
\begin{equation}
\left[\kz^2-\kx^2-\ky^2\right]|k,n\rangle=k(k-1)|k,n\rangle,
\end{equation}
\begin{equation}
\kz|k,n\rangle=(k+n)|k,n\rangle,
\label{eigkz}
\end{equation}
with $n=0,1,\dots,N/2$.  In analogy to Eq.~(\ref{coherento}),
$SU(1,1)$ coherent states are obtained as
\begin{equation} \label{suoocoherento}
|\theta,\phi\rangle \equiv \exp\left(\beta \kp - \beta^*
\km\right)|k,0\rangle~,
\end{equation} 
with $\beta=-(\theta/2)\exp(-i\phi)$ \cite{Barut}.  Power-series
expansion of the exponents in Eq.~(\ref{suoocoherento}) gives the
$SU(1,1)$ coherent states in terms of the number states $|k,n\rangle$,
\begin{equation}
|\theta,\phi\rangle=\left[1-\tanh^2\left({\theta\over
2}\right)\right]^{k}\sum_{n=0}^{N/2}\left[-\tanh\left({\theta\over
2}\right)\exp(-i\phi)\right]^{n}\left(\frac{\Gamma(n+2k)}{n!
\Gamma(2k)}\right)^{1/2}|k,n\rangle~.
\label{suoocoherentt}
\end{equation}
Consequently, the expectation values of $\bf K$ are 
\begin{eqnarray}
\label{kxexp}
\langle\kx\rangle&=&k\sinh\theta\cos\phi,\\
\label{kyexp}
\langle\ky\rangle&=&k\sinh\theta\sin\phi,\\ 
\label{kzexp}
\langle\kz\rangle&=&k\cosh\theta, 
\end{eqnarray}
so that the motion of the vector $\langle {\bf K} \rangle$ is
restricted to the paraboloid
$\langle\kz\rangle^2-\langle\kx\rangle^2-\langle\ky\rangle^2=k^2$ (see
Fig.\ref{fig2}), as could be expected from the $SU(1,1)$ Casimir in
Eq.~(\ref{suoocas}).  The variance of the $SU(1,1)$ generators for the
coherent states (\ref{suoocoherentt}) are given by
\begin{eqnarray}
\label{dkx}
\Delta\kx^2&=&{M\over 2}\left(1+\sinh^2\theta\cos^2\phi\right)~,\\
\label{dky}
\Delta\ky^2&=&{M\over 2}\left(1+\sinh^2\theta\sin^2\phi\right)~,\\
\label{dkz}
\Delta\kz^2&=&{M\over 2}\sinh^2\theta~.
\end{eqnarray}
Due to the possibility of multiple occupation in any single mode,
neither the expectation values nor the variance of the $\bf K$ 
operators are bound. Since the structure constants of the two algebras
differ only in sign, the uncertainty relations of $SU(1,1)$ 
are the same as for $SU(2)$, e.g., 
\begin{equation}
\Delta\kx\Delta\ky\ge {1\over 2}|\langle \kz\rangle|,
\label{kxkyuncert}
\end{equation}
and we can define intelligent and minimum-uncertainty states as we did
for $SU(2)$ in the previous subsection.In Fig \ref{fig2} we plot the
expectation values of $\bf K$ for $SU(2)$ coherent states, as well as
the $\Delta\kx$ and $\Delta\ky$ variance of nine such states.  It is
clear from Eqs.~(\ref{dkx}),(\ref{dky}), and (\ref{kzexp}) that the
intelligent states will be obtained for $\phi=0,\pi/2,\pi,3\pi/2$ and
arbitrary $\theta$.

\subsection{Generalized Squeezing}

As is clear from Eqs.~(\ref{jxjyuncert}) and (\ref{kxkyuncert}), the
minimum fluctuation product of these two observables, depends on the
expectation value of the remaining generator:
\begin{eqnarray}
\Delta X_i^2\le{1\over 2}|c_{ij}^k\langle {\hat X}_k\rangle|, &~{\rm
and} ~& \Delta X_j^2\ge{1\over 2}|c_{ij}^k\langle {\hat X}_k\rangle|.
\end{eqnarray}
Generalized squeezed states of Lie algebras generated by $\{{\hat
X}_i\}_{i=1,2,3}$ are defined as those states for which the variance
in one observable has been reduced at the expense of another
\cite{Wodkiewicz85}.

It is clear from Eqs.~(\ref{jxexp})-(\ref{djz}) that starting from a
fermionic atomic vacuum and inducing a rotation about the $\jx$ axis
(e.g., by choosing $\Delta=0$ and $g=g^*$), the variance in $\jy$ will
be squeezed at the expense of $\jx$ fluctuations (see Fig.~\ref{fig1})
because $1-\sin^2 \theta \le |\cos\theta|$ for all $\theta$.
Similarly, rotating about the $\jy$ axis (e.g., by choosing $\Delta=0$
and $g=-g^*$) will result in generalized squeezing of $\Delta\jx$.
The same is also true for rotations of a boson atomic vacuum, as seen
from Eqs.~(\ref{kxexp})-(\ref{dkz}).  However, the way to attain
generalized squeezing for fermions and bosons is quite different.
Whereas with fermion atoms, squeezing in $\jx$ is obtained by
reduction of its fluctuations, keeping a fixed $\jy$ variance (thereby
reducing the product $\Delta\jx\Delta\jy=|\jz|/2$), the same goal is
attained for boson atoms by {\em increasing} the variance of $\ky$
(thereby increasing $\Delta\kx\Delta\ky=|\kz|/2$)and keeping $\kx$
fluctuations fixed.

\section{Short Time Dynamics}  \label{sect_short_time}

Since the atomic vacuum state (yellow disk in Figs.~\ref{fig1} and
\ref{fig2}) is a coherent state, the atomic states produced in the
dissociation of a molecular BEC will initially be coherent states of
$SU(1,1)$ if the constituent atoms are bosons or of $SU(2)$ when the
constituent atoms are fermions.  There should thus be a significant
difference in the short-time dynamics and in the initial fluctuations
between the two cases.  For bosons, one expects exponential
amplification of atom-number and atom number fluctuations, whereas
fermion number growth is more moderate and fluctuations remain
bound.  Physically, the source of these differences is in the
underlying mechanisms of Bose-stimulation of dissociating boson pairs,
leading to the dynamical instability of the atomic vacuum, and Pauli
blocking of dissociating fermion pairs.

\begin{figure}
\centering
\includegraphics[scale=0.7]{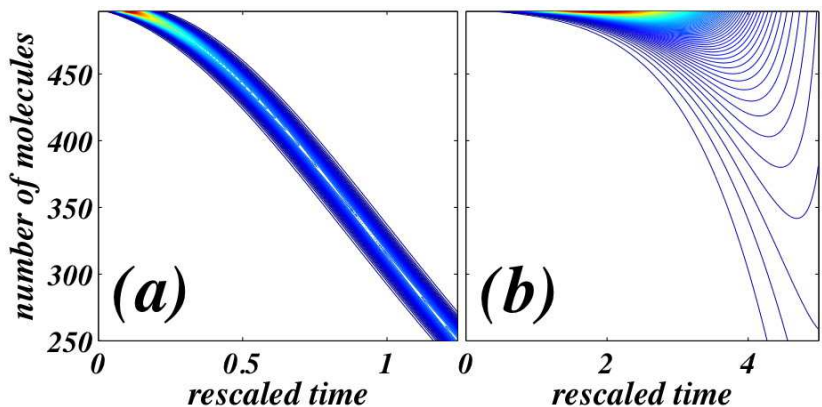}
\caption{(color online) Number distributions as a function of rescaled
time in the dissociation of $500$ molecules into fermion (a) and boson
(b) constituent atoms.}
\label{fig3}
\end{figure}

In order to verify the formation of such coherent states and
generalized squeezing, we have carried out many-particle simulations
of molecular BEC dissociation into either fermionic or bosonic
constituent atoms.  For sufficiently short propagation times the
molecular field is to a good approximation undepleted, and the
generalized operators $\cji,\cki$ coincide, up to an insignificant
$c$-number, with the $SU$ generators $\ji,\ki$.  The atomic states
during the initial stage of dissociation are thus approximately
$SU(2)$ and $SU(1,1)$ coherent states respectively for fermion and
boson constituents.  In what follows, we shall numerically investigate
to what extent do the generalized coherent states and squeezed
fluctuations depicted in section \ref{sect_cs} for an undepleted pump
approximation, carry through to the operators $\cji,\cki$, which
account for pump depletion and fluctuations.  In Fig.~\ref{fig3}, we
plot the atom-number distribution as a function of the rescaled time
$\tau=g\sqrt{N}t$.  The dissociation into fermion constituents shown
in Fig.~\ref{fig3}a, exhibits a Poissonian atom-number distribution
with bound fluctuations as expected from Eqs.~(\ref{jzexp}) and
(\ref{djz}).  Dissociation into boson constituents however, results in
a super-Poissonian number distribution with an exponential growth of
fluctuations as predicted in (\ref{kzexp}) and (\ref{dkz}).

\begin{figure}
\centering
\includegraphics[scale=0.5]{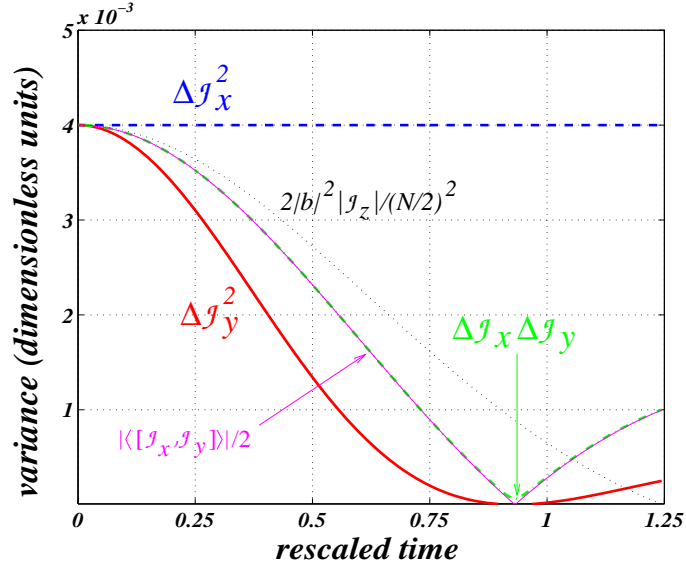}
\caption{(color online) Quadrature variances $\Delta {\cal J}_x^2$
(dashed blue line), $\Delta {\cal J}_y^2$ (solid red line), and
$\Delta {\cal J}_x\Delta {\cal J}_y$ (dash-dotted green line) as a
function of the rescaled time $g\sqrt{N}t$, in the dissociation of a
molecular BEC made of fermion-dimers. Pump phase is $\varphi=0$. The
solid magenta line denotes the uncertainty limit $|\langle\left[{\cal
J}_x,{\cal J}_y\right]\rangle |/2$ which can be approximated at early
times to be $2|b|^2|\cjz|/(N/2)^2$ (dotted black curve).}
\label{fig4}
\end{figure}

\begin{figure}
\centering
\includegraphics[scale=0.5]{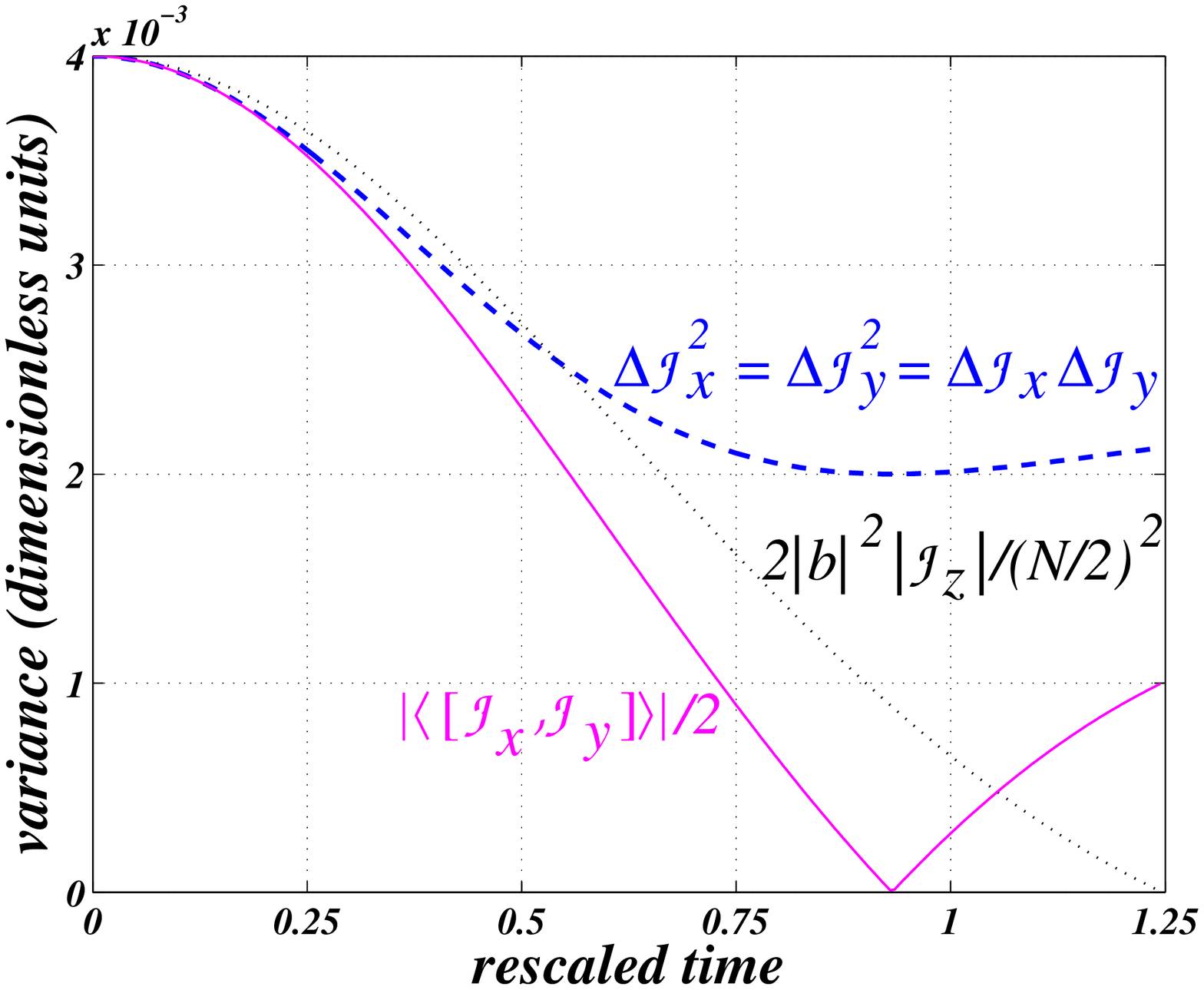}
\caption{(color online) Same as Fig.~\ref{fig3}, with pump phase
$\varphi=\pi/4$. Quadrature variances $\Delta {\cal J}_x^2$, $\Delta
{\cal J}_y^2$, $\Delta {\cal J}_x\Delta {\cal J}_y$ are all equal and
greater than the uncertainty limit $|\langle\left[{\cal J}_x,{\cal
J}_y\right]\rangle |/2$ or its short-time approximation
$2|b|^2|\cjz|/(N/2)^2$ (dotted black curve).}
\label{fig6}
\end{figure}

Generalized $SU(2)$ squeezing and its extension into the depleted-pump
regime is illustrated in Fig.~\ref{fig4} and Fig.~\ref{fig6} where the
variance in $\cjx$ and $\cjy$ in the dynamical evolution of the
(fermion) atomic vacuum state, are plotted as a function of time.  In
Fig.~\ref{fig4}, the phase of the association pump is $\varphi=0$,
corresponding to rotation about the $u$ axis of the Bloch sphere of
Fig.~\ref{fig1}, i.e., $\phi=0$.  The coalescence of the variance
product $\Delta\cjx\Delta\cjy$ with the expectation value of the
commutator $\langle[\cjx,\cjy]\rangle$ demonstrates that the generated
$SU(2)$ coherent state is indeed intelligent.  Squeezing of the
$\Delta\cjy$ variance while keeping a fixed $\Delta\cjx $ is observed,
in agreement with the undepleted pump prediction.  This reduction of
$\cjy$ fluctuations eventually results in the expected minimum
uncertainty state with $\Delta\cjx\Delta\cjy\approx0$.  In comparison,
the evolution of variances for a $\varphi=\pi/4$ phase of the pump
(corresponding to rotation along the $\phi=\pi/4$ circle on the Bloch
sphere of Fig.~\ref{fig1}) is shown in Fig.~\ref{fig6}.  The $SU(2)$
coherent states produced during this evolution, are non-intelligent,
with equal $\Delta\cjx$ and $\Delta\cjy$ variances whose product is
larger than the uncertainty limit.

\begin{figure}
\centering
\includegraphics[scale=0.5]{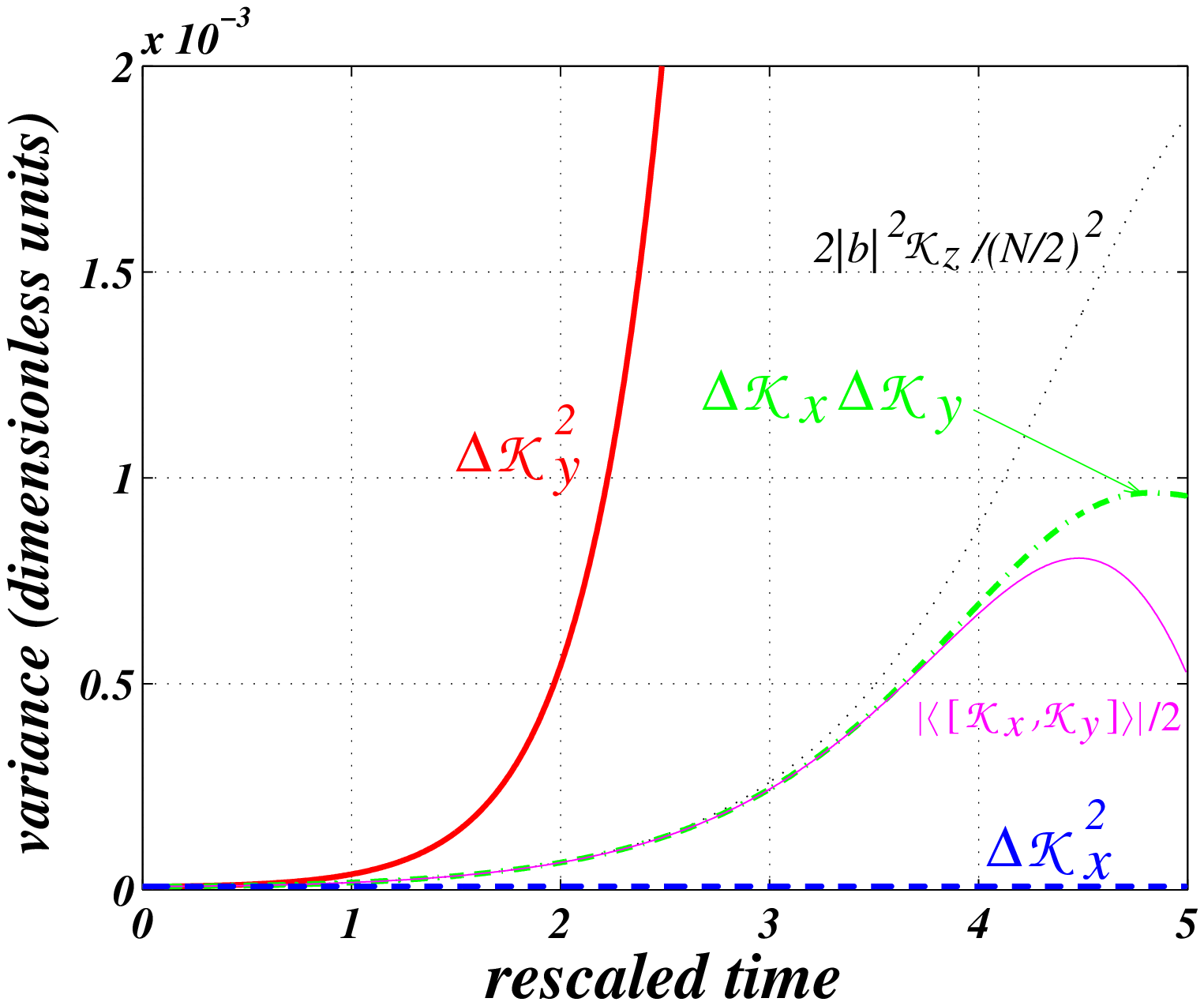}
\caption{(color online) Quadrature variances $\Delta\ckx^2$ (dashed
blue curve), $\Delta\cky^2$ (solid red curve), and
$\Delta\ckx\Delta\cky$ (dash-dotted green curve) as a function of the
rescaled time $g\sqrt{N}t$, in the dissociation of a molecular BEC
made of boson-dimers. Pump phase is $\varphi=0$. The solid magenta
curve denotes the uncertainty limit
$|\langle\left[\ckx,\cky\right]\rangle |/2$ which coincides with
$4|b|^2|\ckz|/(N/2)^2$ (dotted black curve) at early times where
molecular depletion is negligible.}
\label{fig5}
\end{figure}

\begin{figure}
\centering
\includegraphics[scale=0.5]{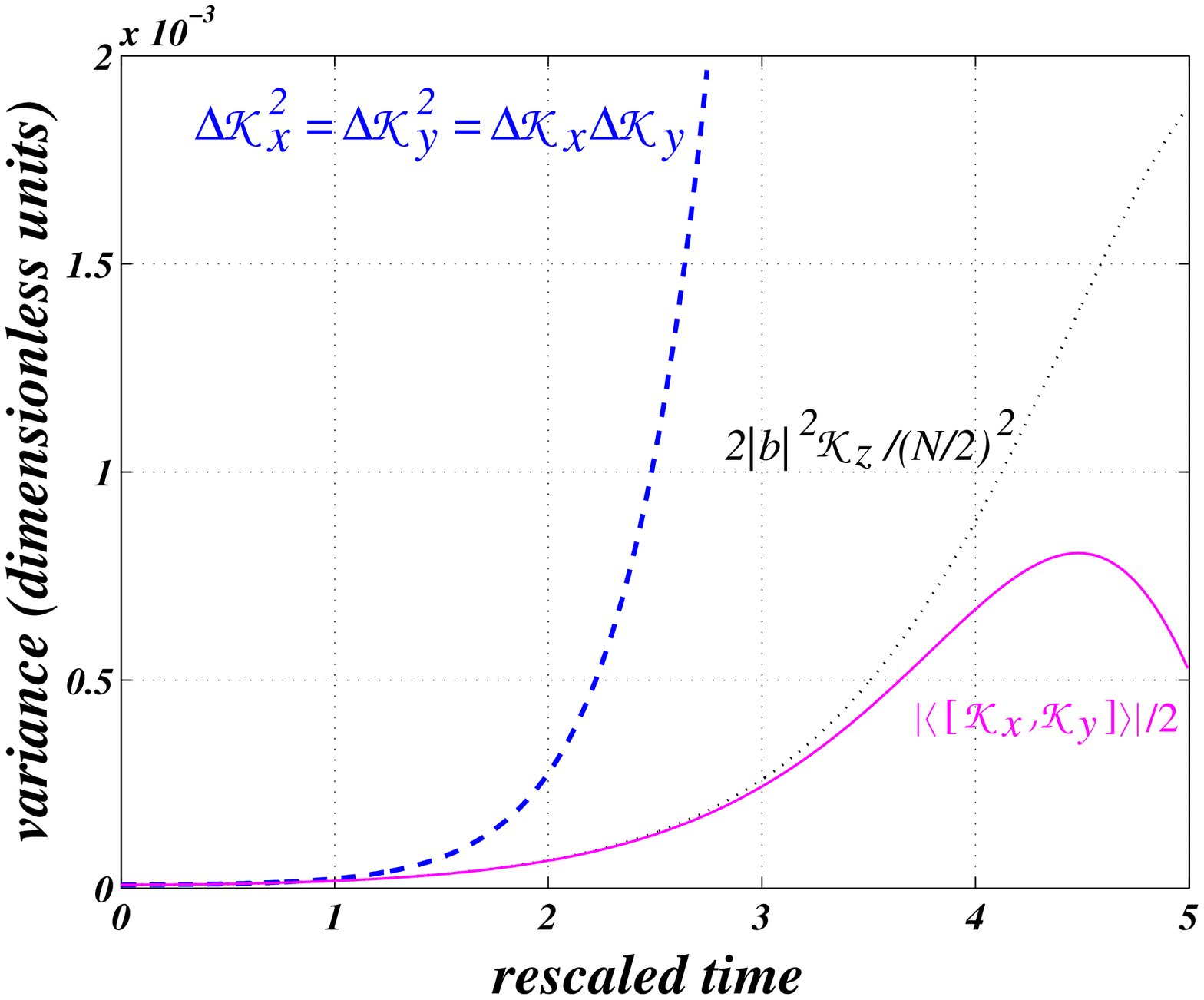}
\caption{(color online) Same as Fig.~\ref{fig4}, with pump phase
$\varphi=\pi/4$. Quadrature variances $\Delta\ckx^2$, $\Delta\cky^2$,
$\Delta\ckx\Delta\cky$ are all equal and greater than the uncertainty
limit $|\langle\left[\ckx,\cky\right]\rangle |/2$ or its short-time
approximation $2|b|^2|\ckz|/(N/2)^2$ (dotted black curve).}
\label{fig7}
\end{figure}

\begin{figure}
\centering
\includegraphics[scale=0.7]{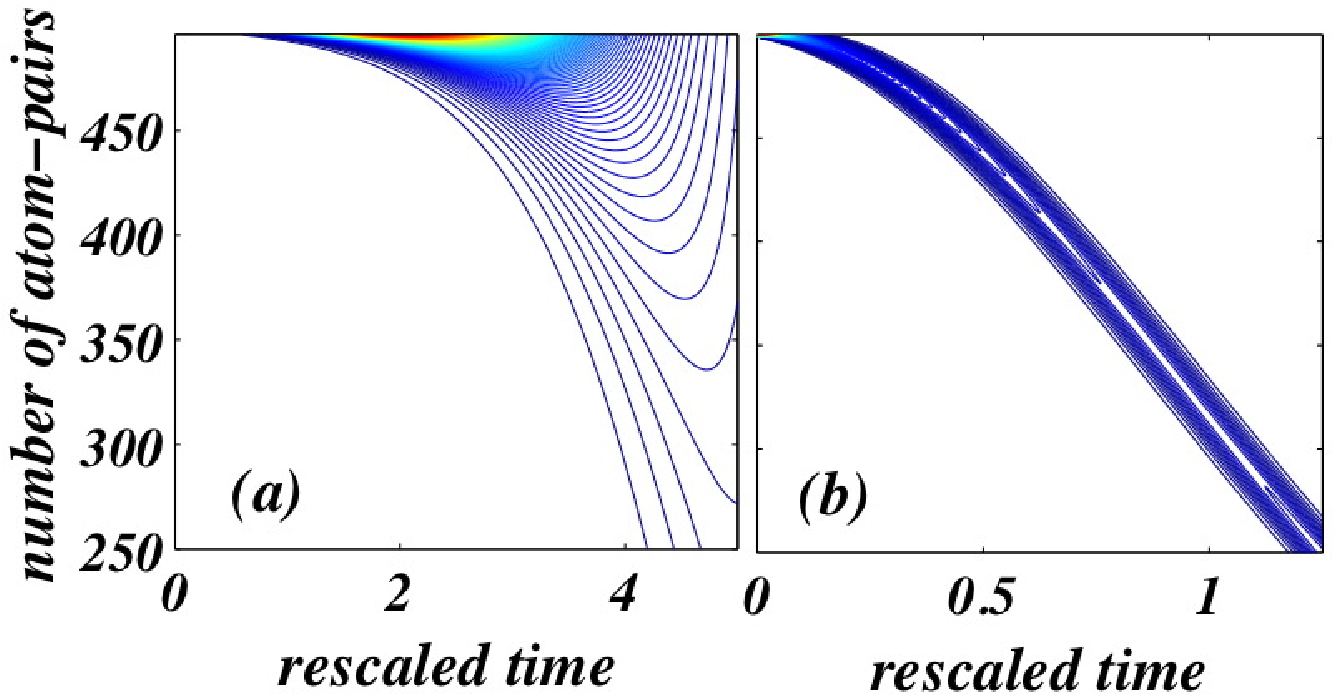}
\caption{(color online) Number distributions as a function of
rescaled time for the association of quantum degenerate gases of 
fermions (a) and bosons (b) with $N=1000$.}
\label{fig8}
\end{figure}

The time evolution of variances in the propagation of the atomic
vacuum state for boson atoms, with $\varphi=0,\pi/4$, is shown
respectively in Fig.~\ref{fig5} and Fig.~\ref{fig7}. Unlike the
fermion case where fluctuations are bound, we observed rapid increase
in the $\Delta\cky$ variance with a fixed $\Delta\ckx$, corresponding
to motion on the $\phi=0$ parabola in Fig.~\ref{fig2}. While the
variance product  $\Delta\ckx\Delta\cky$ grows exponentially with
time, its initial evolution traces the uncertainty limit
$|\langle[\ckx,\cky]\rangle/2$ indicating that the produced states
are indeed $SU(1,1)$ intelligent coherent states. Here too,
propagation with a zero phase of the pump leads to the expected
generalized squeezing. For $\varphi=\pi/4$ however, there is no
squeezing as both $\Delta\ckx$ and $\Delta\cky$ fluctuations are
equal and exponentially growing. The $SU(1,1)$ coherent states
produced are non-intelligent because the variance product
$\Delta\ckx\Delta\cky$ is larger than the uncertainty limit.

The agreement between the numerically-exact variance dynamics of
Figs.~\ref{fig4}-\ref{fig7} and the undepleted pump pictures of
Fig.~\ref{fig1} and Fig.~\ref{fig2}, demonstrates that the different
collective dynamics predicted for fermion and boson constituent atoms
can indeed be interpreted in terms of fluctuations of dynamical
variables quadratic in the atomic creation or annihilation operators.
The reduction of fluctuations of these variables is related to the
formation of coherent states of the $SU(2)$ and $SU(1,1)$ Lie
algebras.  The same qualitative picture seems to apply to the depleted
pump regime.

Finally, the mapping between fermion and boson dynamics, manifested in
the equivalence of Eqs.~(\ref{cjxdot})-(\ref{fersix}) with $\psd=1$
and Eqs.~(\ref{bossix}) with $k=1/2$, is illustrated in
Fig.~\ref{fig8} where atom number distributions are plotted as a
function of time throughout the {\it association} of a Fermi
(Fig.~\ref{fig8}a) and Bose (Fig.~\ref{fig8}b) atomic quantum gas.
Fermion association is mapped onto boson dissociation
(Fig.~\ref{fig3}b) while boson association coincides with fermion
dissociation (Fig.~\ref{fig3}a), generating similar coherent states.

\section{Summary and Conclusions}  \label{sect_summary}

We have established a connection between the collective behavior in
boson and fermion pairing via Feshbach resonances and the generation
of coherent states of the $SU(1,1)$ and $SU(2)$ Lie algebras.  This
relation provides a new viewpoint on the quantum statistics of
atom-molecule quantum gas processes.  The equivalence of molecular BEC
dissociation with the Dicke model for fermion atoms \cite{Dicke54} and
with parametric downconversion for boson atoms \cite{Walls_Milburn},
known for some time in the quantum optics literature, is put into
context as these two quantum systems are paradigmatic examples of the
aforementioned algebras.  The well known squeezing of fluctuations in
dynamical variables linear in the atomic creation or annihilation
operators (e.g. quadrature squeezing) during the dissociation into
boson constituents, may be viewed as generalized $SU(1,1)$ squeezing
of pair fluctuations, quadratic in the atomic creation and
annihilation operators.  Similarly, the coherent evolution during
dissociation into fermion atoms corresponds to the generation of a
minimum uncertainty $SU(2)$ coherent state.  Our numerical simulations
indicate that the same qualitative picture applies to the fluctuation
of operators that account for molecular pump depletion.  The
presentation of the atom-molecule system in terms of these generators
offers a link between the fermion- and boson-constituent atom cases
due to the close relation and direct mapping between the underlying
Lie algebras.

\begin{acknowledgments}
This work was supported in part by grants from the 
Minerva foundation for a junior research group, the
Israel Science Foundation (Center of Excellence grant No.~8006/03 and
Personal grant Nos.~582/07,~29/07), the U.S.-Israel
Binational Science Foundation (grant Nos.~2002147 and 2006212),  
and the James Franck German-Israeli Binational Program in Laser-Matter 
Interactions.
\end{acknowledgments}

\end{document}